\documentclass[sigconf]{acmart}

\usepackage{multirow}
\usepackage[capitalise]{cleveref}
\usepackage{makecell}
\usepackage{balance}

\AtBeginDocument{%
  }
\copyrightyear{2026}
\acmYear{2026}
\setcopyright{cc}
\setcctype{by-nc-nd}
\acmConference[UMAP '26]{34th ACM Conference on User Modeling, Adaptation and Personalization}{June 08--11, 2026}{Gothenburg, Sweden}
\acmBooktitle{34th ACM Conference on User Modeling, Adaptation and Personalization (UMAP '26), June 08--11, 2026, Gothenburg, Sweden}
\acmDOI{10.1145/3774935.3806149}
\acmISBN{979-8-4007-2311-7/2026/06}

\begin{document}

\title[Trust as a Situated User State]{Trust as a Situated User State in Social LLM-Based Chatbots}
\subtitle{A Longitudinal Study of Snapchat's My AI}
%

\author{Annie Landerberg}
\authornote{Equal contribution.}
\orcid{0009-0000-8792-3374}
\affiliation{%
  \institution{University of Gothenburg}
  \city{Gothenburg}
  \country{Sweden}
}
\email{annie.landerberg@hotmail.com}

\author{Kari Flatmo}
\authornotemark[1]
\orcid{0009-0008-8181-5450}
\affiliation{%
  \institution{University of Gothenburg}
  \city{Gothenburg}
  \country{Sweden}
}
\email{kari.flatmo@hotmail.com}

\author{Alan Said}
\orcid{0000-0002-2929-0529}
\affiliation{%
  \institution{University of Gothenburg}
  \city{Gothenburg}
  \country{Sweden}}
\email{alansaid@acm.org}

\renewcommand{\shortauthors}{Landerberg, Flatmo, et al.}

\begin{abstract}
Social chatbots based on large language models are increasingly embedded in everyday platforms, yet how users develop trust in these systems over time remains unclear. We present a four-week longitudinal qualitative survey study (N = 27) of trust formation in Snapchat's My AI, a socially embedded conversational agent. Our findings show that trust is shaped by perceived ability, conversational behavior, human-likeness, transparency, privacy concerns, and trust in the host platform. Trust does not remain stable, but evolves through interaction as users adapt their expectations, refine their prompting strategies, and actively regulate how and when they rely on the system. These processes reflect a continuous negotiation of trust, not a one-time evaluation. While conversational fluency supports engagement, excessive anthropomorphism and limited transparency can undermine trust over time. We synthesize these findings into a conceptual model that frames trust as a dynamic user state shaped by interaction context and expectations, with implications for the design of human-centered and adaptive conversational agents.
\end{abstract}



\maketitle

\section{Introduction}
Large language model (LLM)–based conversational agents are increasingly integrated into everyday digital platforms, enabling more naturalistic and socially oriented forms of human–AI interaction. Beyond task-oriented applications such as customer service or information retrieval, social chatbots are designed to act as conversational partners, companions, or sources of social support, engaging users through human-like language, affective cues, and continuous availability. As these systems become embedded in personal communication environments, questions of user trust become central to their adoption, use, and long-term impact.

Trust has long been recognized as a critical factor in HCI and automation, shaping users’ willingness to rely on, disclose information to, and engage with intelligent systems \cite{mayer1995,lee_trust_2004,mcknight_trust_2011}. Prior work on trust in conversational agents has primarily focused on task--oriented contexts, such as customer service chatbots, where trust is linked to correctness, efficiency, and perceived expertise \cite{Folstad2018WhatMU,nordheim2019initial}. In parallel, work on social chatbots and AI companions has shown that human-like interaction styles can foster engagement, self-disclosure, and relational bonding, while also introducing new risks related to privacy, emotional dependence, and miscalibrated trust \cite{skjuve_my_2021,brandtzaeg2022my}.

Despite this growing body of work, comparatively little is known about how users establish and negotiate trust in social chatbots that are embedded within commercial social media platforms and powered by contemporary LLM technologies. These systems differ from earlier conversational agents in both capability and context. They operate within existing platforms that already carry trust, privacy, and economic expectations, and they blur boundaries between social interaction, information provision, and platform-level data practices. Understanding trust in such settings requires attention not only to the chatbot’s conversational behavior, but also to users’ expectations, perceptions of risk, and the broader service environment in which the interaction takes place.

To address this gap, we investigate users’ experiences of trust in Snapchat’s My AI, a widely deployed social chatbot integrated into the Snapchat messaging platform and powered by GPT-based language models. Drawing on a four-week longitudinal qualitative survey study with 27 young adult users, we examine how trust is shaped by chatbot-related factors such as perceived ability, conversational behavior, and human-likeness, as well as environment-related factors including transparency, privacy, perceived risk, and trust in the host platform itself.

Building on established models of interpersonal trust \cite{mayer1995} and prior work on trust in chatbots \cite{nordheim2019initial}, this paper conceptualizes trust as a situated and evolving user experience, not a static system property. From a user modeling perspective, this suggests that trust can be understood as a latent user state that evolves through interaction and shapes how users interpret, engage with, and rely on conversational agents over time.

Rather than identifying new trust factors, the contribution lies in showing how these factors are negotiated and reshaped through interaction with social chatbots over time.


\section{Related Work}
\label{sec:relwork}
Trust is a central concept in human-AI interaction, influencing users’ willingness to rely on, engage with, and disclose information to intelligent systems \cite{mayer1995,lee_trust_2004,mcknight_trust_2011}. In AI-mediated interaction, trust depends not only on system performance but also on perceptions of transparency, intent, and risk \cite{hoff2015,glikson2020}.

Within conversational agent research, trust has primarily been studied in task-oriented contexts such as customer service. \citet{Folstad2018WhatMU} identify correct interpretation, professional presentation, and perceived expertise as key drivers of trust, while \citet{nordheim2019initial} distinguish between chatbot-related, environment-related, and user-related factors. However, this work largely focuses on instrumental interactions, leaving socially oriented use less explored.

Research on social chatbots and AI companions emphasizes relational and affective interaction. These systems are designed to support companionship, emotional expression, and long-term engagement \cite{zhou_design_2020,ta2020}. Human-like conversational behavior and perceived empathy can foster engagement and self-disclosure \cite{skjuve_my_2021,brandtzaeg2022my}, but may also introduce risks such as emotional dependence \cite{rodogno_social_2016,laestadius_too_2024}. More generally, anthropomorphic cues can increase social presence and trust \cite{ciechanowski2019shades,de_cicco_millennials_2020}, but excessive human-likeness may reduce trust due to discomfort or uncanny experiences \cite{muresanChatsBotsBalancing2019,skjuve_help_2019}.

Transparency, privacy, and data use are also central. Uncertainty about how personal data are handled can undermine trust and willingness to engage \cite{zamora_im_2017,ischen_privacy_2020}, while trust in the organization behind the chatbot shapes perceived risk and acceptability \cite{Folstad2018WhatMU,brandtzaeg2019losing}.

Despite this work, there remains limited empirical research examining how trust develops over time in social chatbots embedded in commercial platforms and powered by contemporary large language models. This paper addresses this gap by examining trust as a situated and evolving user experience shaped by conversational behavior, user expectations, and platform context.

\section{Method}
We conducted a longitudinal qualitative survey to investigate how users experience and establish trust when interacting with a social chatbot in everyday use. The study focused on \emph{Snapchat’s My AI}, a GPT-based conversational agent in Snapchat's messaging platform. The system was selected due to its widespread use and integration into everyday messaging practices.

\subsection{Participants}
Participants were recruited using convenience and snowball sampling through the authors’ personal and professional networks. Inclusion criteria required participants to be between 18 and 30 years old, be active Snapchat users, and have limited prior experience interacting with Snapchat’s My AI. A total of 27 participants took part in the study. Participants were Swedish or Norwegian, reflecting Snapchat’s widespread use among young adults in these countries \cite{IpsosSoMeTrackerQ4232024}. 
Of the 27 participants, 25 completed all three survey rounds.
\cref{tab:participant-demographics} presents the demographic details of the participants.

\begin{table}
\caption{Participant demographics by age group.}
\label{tab:participant-demographics}

\centering
\footnotesize
\begin{tabular}{|r|r|r|r|}
\hline

\textbf{Age group} & \textbf{Female} & \textbf{Male} & \textbf{Total}  \\
\hline
20-23 &  5 & 2 & 7 \\
24-27 &  8 & 7 & 15\\
28-30 &  3 & 2 & 5\\
\hline
\end{tabular}
\end{table}

\subsection{Study Design and Procedure}
The study was conducted over a four-week period in 2024 during which participants were asked to interact freely with My AI as part of their everyday Snapchat use. Data were collected through three online qualitative surveys administered at regular intervals: after one week, mid-way through the study, and at the end of the four-week period. This design enabled exploration of both initial impressions and changes in perceptions over time.

The surveys consisted of open-ended questions derived from prior literature on trust in chatbots and interpersonal trust\footnote{The questions are available in the paper's supplemental material.}. Questions addressed chatbot-related factors such as perceived ability, conversational behavior, and human-likeness, as well as environment-related factors including transparency, privacy, perceived risk, and the role of the Snapchat platform. Participants could respond in Swedish, Norwegian, or English.

Participants were not given predefined tasks, but were encouraged to integrate My AI into their everyday use. This allowed the study to capture a range of naturalistic interactions without restricting use to specific scenarios. At the same time, the survey questions prompted participants to reflect on concrete interactions, including how they used the system, how they evaluated its responses, and how their trust changed over time. The three survey rounds captured initial impressions, emerging usage patterns, and later reflections after sustained use, enabling exploration of trust formation while maintaining consistency across responses.

\subsection{Data Analysis}
Across the three surveys, a total of 1,126 responses were analyzed. Following reflexive thematic analysis \cite{braun2006}, two of the authors familiarized themselves with the data and developed initial codes, which were discussed and refined collaboratively. Themes were then developed and defined through comparison across participants and survey rounds. In line with reflexive thematic analysis, coding was treated as an interpretive process and inter-rater reliability metrics were not applied.

Responses from individual participants were examined longitudinally to identify changes in perceptions and experiences of trust over time. To support transparency regarding prevalence, themes are described using approximate frequency terms, with approximate proportions (e.g., ``a few'' $<25\%$, ``some'' $25–50\%$, ``most'' $>50\%$) used to indicate prevalence.

The analysis moved iteratively between coding and theme development, with codes refined as patterns became clearer. Particular attention was given to how responses changed over time within participants, enabling identification of different trajectories of trust development. Themes were reviewed in relation to both individual responses and the full dataset to ensure consistency and coherence.

\subsection{Ethical Considerations}
All participants received information about the study’s purpose and provided informed consent prior to participation. Participants were informed of their right to withdraw at any time. Personal identifiers were removed prior to analysis. Pseudonyms were used when reporting illustrative quotes. The study complied with applicable ethical guidelines for research involving human participants.



\section{Findings}
This section presents findings from the three qualitative surveys conducted over four weeks, focusing on factors that shaped participants’ establishment of trust in Snapchat’s My AI. The results are organized into chatbot-related and environment-related factors, followed by changes in perceptions over time. These factors describe dimensions along which users’ trust-related perceptions varied, and which may be relevant for modeling user trust dynamics in socially embedded conversational agents, see \cref{tab:themes} for an overview.

\begin{table}
\caption{Dimensions of user trust in a socially embedded conversational agent.}
\label{tab:themes}
\centering
\begin{tabular}{|p{1.2cm}|p{2cm}|p{4.3cm}|}
\hline
\textbf{Category} & \textbf{Factor} & \textbf{Factor details} \\
\hline
 & Ability and expertise & Users’ perception of My AI’s expertise; ability to provide correct and useful information. \\
\cline{2-3}
\multirow{3}{*}{Chatbot} & Conversation and language & Experience of conversation style; ability to interpret and understand user language. \\
\cline{2-3}
 & Human-likeness & Perception of human-likeness and ``empathy''. \\
 
\hline
\multirow{3}{*}{\makecell{Environ-\\ment}} & Transparency and privacy & Perception of transparency and handling of personal data. \\
\cline{2-3}
 & Perceived risk & Risks associated with using My AI. \\
\cline{2-3}
 & The Snapchat platform & How integration in Snapchat affects use and perception. \\
\hline
\end{tabular}
\end{table}

\subsection{Chatbot-Related Factors}

\emph{Ability and expertise} played a central role in shaping participants’ trust. Most participants experienced My AI as capable of providing helpful and relevant responses for everyday use, such as recommendations, creative inspiration, or general advice. Several participants reported following the chatbot’s suggestions in low-stakes contexts. As one participant noted, ``Surprisingly good answers. I could make a banger carbonara from what it told me'' (Benjamin, Survey 1).

At the same time, trust in the chatbot’s factual correctness was often limited. Many participants expressed reluctance to rely on My AI for complex, scientific, or verifiable information, citing the absence of sources and transparency. One participant stated, ``You can’t verify that what she is saying is correct. I wouldn’t use it for any other purpose than inspiration or ideas'' (Fred, Survey 2). These perceptions indicate that perceived usefulness did not necessarily translate into full trust.

Expectations strongly influenced how ability and expertise were evaluated. Participants who approached My AI as an information-seeking or search-oriented tool often expressed disappointment when responses failed to meet expectations shaped by prior experience with other AI systems. In contrast, participants who framed My AI as a social chatbot were more tolerant of imprecision and evaluated its performance more positively.

\emph{Conversation and language} were generally perceived as strengths of My AI. Most participants appreciated the chatbot’s quick response time, ability to interpret questions, and use of follow-up prompts, which lowered the threshold for interaction. Its conversational style was often described as friendly and approachable. However, some participants experienced the interaction as shallow or overly generic. One participant described the chatbot as ``pretending to be your friend but trying to figure you out, but not very subtle'' (Fredrika, Survey 1).

\emph{Human-likeness} emerged as a double-edged factor for trust. Many participants described My AI as polite and friendly, and several reported instinctively using greetings or polite expressions despite knowing they were interacting with a chatbot. As one participant explained, ``Because the way it responds is similar enough to how a human would speak, it is almost instinct to be polite'' (Fredrika, Survey 1). These human-like cues made interaction feel natural and encouraged engagement.

At the same time, excessive friendliness or perceived empathy was described by some participants as uncomfortable or inauthentic. One participant reflected that the chatbot felt ``like a system to stroke one’s ego rather than something constructive'' (Vidar, Survey 2). Such reactions often led to reduced trust and reluctance to continue interacting with the chatbot.

\subsection{Environment-Related Factors}
Across all three surveys, \emph{transparency and privacy} concerns were among the most salient factors affecting trust. Nearly all participants expressed uncertainty about what data My AI collected, whether conversations were stored, and how information might be used by Snapchat. This uncertainty led many participants to actively limit what they shared. One participant stated, ``It’s scary not to know what’s being done with the information and lose control over your own information'' (Karolina, Survey 1).

\emph{Perceived risk} was closely tied to these concerns. Participants mentioned risks related to data misuse, targeted advertising, and the broader implications of interacting with a rapidly developing AI technology. Some participants also raised concerns about emotional dependence, particularly for vulnerable users. As one participant noted, ``I cannot see the long-term benefit of being ‘friends’ with something that doesn’t exist'' (Albin, Survey 1).

The \emph{Snapchat platform} itself emerged as an important contextual factor influencing trust. For some participants, My AI’s integration into a familiar platform lowered barriers to use and made interaction feel natural. For others, Snapchat’s reputation as a commercial social media platform heightened concerns about profit-driven motives and data exploitation. One participant reflected, ``It keeps sending me adverts. It’s shoving in my face that it’s all about earning money'' (Fredrika, Survey 1).

These factors were not independent, but were often evaluated together, with, for example, perceived ability influencing how transparency and risk were interpreted.

\subsection{Changes Over Time}
Participants’ experiences of My AI evolved over the four-week period. Some participants reported that interaction became more natural over time and that they discovered new, practical uses for the chatbot. As one participant noted, ``It’s become more natural after a month to ask it questions and that the threshold for talking to it is lower'' (Julia, Survey 3).

Others became increasingly critical as they noticed limitations related to transparency, accuracy, or commercial content. No participant described developing a strong social bond with the chatbot, and several explicitly stated that they would not continue using My AI after the study due to trust-related concerns.

Overall, the findings show that trust in a socially embedded chatbot is dynamic and shaped by ongoing interaction, evolving expectations, and the broader platform context of the chatbot.

\section{Discussion}
This study examined how users establish trust in a socially embedded, LLM-based chatbot through everyday interaction over time. Focusing on Snapchat’s My AI, the findings extend prior work on trust in conversational agents by showing that trust is a situated and evolving user experience shaped jointly by interactional behavior, user expectations, and platform context.

Consistent with prior research, perceived ability and expertise were central to trust evaluations \cite{mayer1995,Folstad2018WhatMU,nordheim2019initial}. The findings show that expertise was assessed relative to users’ expectations, not as an absolute system property. Participants who approached My AI as an information or search tool often evaluated it against standards shaped by prior experience with other AI systems, leading to skepticism. Participants who framed My AI as a social chatbot were more tolerant of imprecision and evaluated its performance more positively. This highlights the role of expectation alignment in shaping trust in social conversational agents.

Human-likeness further influenced trust in ambivalent ways. Politeness, friendliness, and conversational fluency supported engagement and lowered interaction barriers, often prompting users to apply social norms despite knowing they were interacting with a chatbot. At the same time, excessive friendliness or perceived empathy reduced trust for some participants, who described such behavior as uncomfortable or inauthentic. This supports prior findings that trust depends on careful calibration of anthropomorphic cues, not their maximization \cite{ciechanowski2019shades,muresanChatsBotsBalancing2019,skjuve_help_2019}.

Transparency and privacy concerns emerged as persistent barriers to trust. Even when participants found My AI useful, uncertainty about data collection and use limited their willingness to disclose information or rely on the chatbot. These concerns were closely tied to perceptions of Snapchat as a commercial platform, extending prior work on service context and brand trust by showing that trust in socially embedded chatbots is distributed across both the agent and its host platform \cite{Folstad2018WhatMU,brandtzaeg2019losing}.

Across the four-week period, trust did not converge toward acceptance or rejection. Participants’ perceptions evolved in different directions depending on how interactions developed and which concerns became more prominent. Some participants reported increased familiarity and lowered interaction thresholds, while others became more critical over time. These findings suggest that trust cannot be treated as a fixed user attribute, but as a temporally evolving state that interacts with personalization, expectation management, and interface design.

Taken together, the findings suggest that designing trustworthy social chatbots requires attention not only to conversational competence, but also to expectation management, calibrated human-likeness, and transparency within the broader platform context. For example, systems could make uncertainty and data usage more explicit, or support user verification practices directly in the interface. While this study does not evaluate adaptive mechanisms directly, it highlights trust as a key experiential dimension shaping user engagement with socially embedded conversational agents.

\section{Conclusion}
This work synthesizes insights from a four-week longitudinal qualitative study of trust in Snapchat’s My AI, a socially embedded conversational agent. The findings show that users’ trust is shaped by a combination of chatbot-related factors, including perceived ability, conversational behavior, and human-likeness, as well as environment-related factors such as transparency, privacy concerns, and trust in the host platform. Rather than converging toward stable acceptance or rejection, trust emerged as a dynamic and context-sensitive user experience evolving with continued use.

The study highlights that conversational fluency and human-like behavior can support engagement without necessarily producing trust, particularly when transparency and expectation alignment are lacking. These findings highlight the importance of considering trust not as a static system property, but as an experiential user state that evolves through interaction and may inform the design of adaptive and personalized conversational systems. For human-centered conversational agents, this suggests that managing expectations, calibrating human-likeness, and communicating limitations may be as important as improving language generation capabilities.

This work has several limitations. First, the study relies on a qualitative sample of young adult users from two countries, which may limit generalizability to other populations or contexts. Second, the four-week study duration captures early and short-term dynamics of trust but cannot account for longer-term use or sustained relationships with social chatbots. Third, the findings are specific to Snapchat’s My AI and its platform integration, and may not transfer directly to other social or task-oriented conversational agents.

Despite these limitations, the study provides empirical insight into trust formation in socially embedded conversational agents and contributes to ongoing discussions on human-centered AI. Understanding how trust develops in real-world contexts is a necessary step toward designing adaptive systems that support appropriate reliance, engagement, and user control.

\begin{acks}
Generative AI was used in typesetting, phrasing, spellchecking, and grammar correction of this text.
\end{acks}
\balance

\bibliographystyle{ACM-Reference-Format}
\bibliography{bibliography}

@article{ta2020,
author="Ta, Vivian
and Griffith, Caroline
and Boatfield, Carolynn
and Wang, Xinyu
and Civitello, Maria
and Bader, Haley
and DeCero, Esther
and Loggarakis, Alexia",
title="User Experiences of Social Support From Companion Chatbots in Everyday Contexts: Thematic Analysis",
journal="J Med Internet Res",
year="2020",
month="Mar",
day="6",
volume="22",
number="3",
pages="e16235",
issn="1438-8871",
doi="10.2196/16235",
url="http://www.jmir.org/2020/2/e16235/",
url="https://doi.org/10.2196/16235",
url="http://www.ncbi.nlm.nih.gov/pubmed/32141837"
}

@article{mayer1995,
	title = {An {Integrative} {Model} of {Organizational} {Trust}},
	volume = {20},
	issn = {03637425},
	doi = {10.2307/258792},
	number = {3},
	urldate = {2022-08-10},
	journal = {The Academy of Management Review},
	author = {Mayer, Roger C. and Davis, James H. and Schoorman, F. David},
	month = jul,
	year = {1995},
	note = {Publisher: Academy of Management},
	pages = {709},
}

@article{hoff2015,
author = {Kevin Anthony Hoff and Masooda Bashir},
title ={Trust in Automation: Integrating Empirical Evidence on Factors That Influence Trust},
journal = {Human Factors},
volume = {57},
number = {3},
pages = {407-434},
year = {2015},
doi = {10.1177/0018720814547570},
note ={PMID: 25875432},
URL = {https://doi.org/10.1177/0018720814547570},
eprint = {https://doi.org/10.1177/0018720814547570},
}

@article{glikson2020,
	title = {Human {Trust} in {Artificial} {Intelligence}: {Review} of {Empirical} {Research}},
	volume = {14},
	issn = {1941-6520},
	shorttitle = {Human {Trust} in {Artificial} {Intelligence}},
	url = {https://journals.aom.org/doi/10.5465/annals.2018.0057},
	doi = {10.5465/annals.2018.0057},
	number = {2},
	urldate = {2024-06-05},
	journal = {Academy of Management Annals},
	author = {Glikson, Ella and Woolley, Anita Williams},
	month = jul,
	year = {2020},
	note = {Publisher: Academy of Management},
	pages = {627--660},
}

@inproceedings{muresanChatsBotsBalancing2019,
  title = {Chats with {{Bots}}: {{Balancing Imitation}} and {{Engagement}}},
  shorttitle = {Chats with {{Bots}}},
  booktitle = {Extended {{Abstracts}} of the 2019 {{CHI Conference}} on {{Human Factors}} in {{Computing Systems}}},
  author = {Muresan, Andreea and Pohl, Henning},
  year = {2019},
  month = may,
  series = {{{CHI EA}} '19},
  pages = {1--6},
  publisher = {Association for Computing Machinery},
  address = {New York, NY, USA},
  doi = {10.1145/3290607.3313084},
  urldate = {2025-05-09},
  isbn = {978-1-4503-5971-9}
}

@article{zhou_design_2020,
	title = {The {Design} and {Implementation} of {XiaoIce}, an {Empathetic} {Social} {Chatbot}},
	volume = {46},
	issn = {0891-2017},
	url = {https://doi.org/10.1162/coli_a_00368},
	doi = {10.1162/coli_a_00368},
	number = {1},
	urldate = {2024-06-05},
	journal = {Computational Linguistics},
	author = {Zhou, Li and Gao, Jianfeng and Li, Di and Shum, Heung-Yeung},
	month = mar,
	year = {2020},
	pages = {53--93},
}

@article{nordheim2019initial,
	title = {An initial model of trust in chatbots for customer service—{Findings} from a questionnaire study},
	volume = {31},
	issn = {1873-7951},
	doi = {10.1093/iwc/iwz022},
	number = {3},
	journal = {Interacting with Computers},
	author = {Nordheim, Cecilie Bertinussen and Følstad, Asbjørn and Bjørkli, Cato Alexander},
	year = {2019},
	note = {Place: United Kingdom
Publisher: Oxford University Press},
	pages = {317--335},
}

@inproceedings{Folstad2018WhatMU,
	address = {Cham},
	title = {What {Makes} {Users} {Trust} a {Chatbot} for {Customer} {Service}? {An} {Exploratory} {Interview} {Study}},
	isbn = {978-3-030-01437-7},
	shorttitle = {What {Makes} {Users} {Trust} a {Chatbot} for {Customer} {Service}?},
	doi = {10.1007/978-3-030-01437-7_16},
	language = {en},
	booktitle = {Internet {Science}},
	publisher = {Springer International Publishing},
	author = {Følstad, Asbjørn and Nordheim, Cecilie Bertinussen and Bjørkli, Cato Alexander},
	editor = {Bodrunova, Svetlana S.},
	year = {2018},
	pages = {194--208},
}

@article{ciechanowski2019shades,
title = {In the shades of the uncanny valley: An experimental study of human–chatbot interaction},
journal = {Future Generation Computer Systems},
volume = {92},
pages = {539-548},
year = {2019},
issn = {0167-739X},
doi = {https://doi.org/10.1016/j.future.2018.01.055},
url = {https://www.sciencedirect.com/science/article/pii/S0167739X17312268},
author = {Leon Ciechanowski and Aleksandra Przegalinska and Mikolaj Magnuski and Peter Gloor},
}

@article{brandtzaeg2022my,
  title = {My {{AI}} Friend: {{How}} Users of a Social Chatbot Understand Their Human--{{AI}} Friendship},
  shorttitle = {My {{AI}} Friend},
  author = {Brandtzaeg, Petter Bae and Skjuve, Marita and F{\o}lstad, Asbj{\o}rn},
  year = {2022},
  journal = {Human Communication Research},
  volume = {48},
  number = {3},
  pages = {404--429},
  publisher = {Oxford University Press},
  address = {United Kingdom},
  issn = {1468-2958},
  doi = {10.1093/hcr/hqac008}
}

@article{brandtzaeg2019losing,
  title = {Losing {{Control}} to {{Data-Hungry Apps}}: {{A Mixed-Methods Approach}} to {{Mobile App Privacy}}},
  shorttitle = {Losing {{Control}} to {{Data-Hungry Apps}}},
  author = {Brandtzaeg, Petter Bae and Pultier, Antoine and Moen, Gro Mette},
  year = {2019},
  month = aug,
  journal = {Social Science Computer Review},
  volume = {37},
  number = {4},
  pages = {466--488},
  publisher = {SAGE Publications Inc},
  issn = {0894-4393},
  doi = {10.1177/0894439318777706},
  urldate = {2025-05-09},
  langid = {english}
}

@article{braun2006,
  title = {Using Thematic Analysis in Psychology},
  author = {Braun, Virginia and {and Clarke}, Victoria},
  year = {2006},
  month = jan,
  journal = {Qualitative Research in Psychology},
  volume = {3},
  number = {2},
  pages = {77--101},
  publisher = {Routledge},
  issn = {1478-0887},
  doi = {10.1191/1478088706qp063oa},
  urldate = {2025-05-09}
}

@inproceedings{ischen_privacy_2020,
    address = {Cham},
    title = {Privacy {Concerns} in {Chatbot} {Interactions}},
    isbn = {978-3-030-39540-7},
    doi = {10.1007/978-3-030-39540-7_3},
    language = {en},
    booktitle = {Chatbot {Research} and {Design}},
    publisher = {Springer International Publishing},
    author = {Ischen, Carolin and Araujo, Theo and Voorveld, Hilde and van Noort, Guda and Smit, Edith},
    editor = {Følstad, Asbjørn and Araujo, Theo and Papadopoulos, Symeon and Law, Effie Lai-Chong and Granmo, Ole-Christoffer and Luger, Ewa and Brandtzaeg, Petter Bae},
    year = {2020},
    pages = {34--48},
}

@article{lee_trust_2004,
    title = {Trust in {Automation}: {Designing} for {Appropriate} {Reliance}},
    volume = {46},
    issn = {0018-7208},
    shorttitle = {Trust in {Automation}},
    url = {https://journals.sagepub.com/action/showAbstract},
    doi = {10.1518/hfes.46.1.50_30392},
    language = {EN},
    number = {1},
    urldate = {2026-01-28},
    journal = {Human Factors},
    author = {Lee, John D. and See, Katrina A.},
    month = mar,
    year = {2004},
    note = {Publisher: SAGE Publications Inc},
    pages = {50--80},
}

@article{mcknight_trust_2011,
    title = {Trust in a specific technology: {An} investigation of its components and measures},
    volume = {2},
    issn = {2158-656X},
    shorttitle = {Trust in a specific technology},
    url = {https://dl.acm.org/doi/10.1145/1985347.1985353},
    doi = {10.1145/1985347.1985353},
    number = {2},
    urldate = {2026-01-28},
    journal = {ACM Trans. Manage. Inf. Syst.},
    author = {Mcknight, D. Harrison and Carter, Michelle and Thatcher, Jason Bennett and Clay, Paul F.},
    month = jul,
    year = {2011},
    pages = {12:1--12:25},
}

@article{skjuve_my_2021,
    title = {My {Chatbot} {Companion} – a {Study} of {Human}-{Chatbot} {Relationships}},
    copyright = {Navngivelse 4.0 Internasjonal},
    issn = {1071-5819},
    url = {https://sintef.brage.unit.no/sintef-xmlui/handle/11250/3048748},
    doi = {10.1016/j.ijhcs.2021.102601},
    language = {eng},
    urldate = {2024-06-06},
    journal = {14},
    author = {Skjuve, Marita and Følstad, Asbjørn and Fostervold, Knut Inge and Brandtzæg, Petter Bae},
    year = {2021},
    note = {Accepted: 2023-02-07T08:42:43Z
Publisher: Elsevier},
}

@article{rodogno_social_2016,
    title = {Social robots, fiction, and sentimentality},
    volume = {18},
    issn = {1572-8439},
    url = {https://doi.org/10.1007/s10676-015-9371-z},
    doi = {10.1007/s10676-015-9371-z},
    language = {en},
    number = {4},
    urldate = {2026-01-28},
    journal = {Ethics and Information Technology},
    author = {Rodogno, Raffaele},
    month = dec,
    year = {2016},
    pages = {257--268},
}

@article{laestadius_too_2024,
    title = {Too human and not human enough: {A} grounded theory analysis of mental health harms from emotional dependence on the social chatbot {Replika}},
    volume = {26},
    issn = {1461-4448},
    shorttitle = {Too human and not human enough},
    url = {https://doi.org/10.1177/14614448221142007},
    doi = {10.1177/14614448221142007},
    language = {EN},
    number = {10},
    urldate = {2026-01-28},
    journal = {New Media \& Society},
    author = {Laestadius, Linnea and Bishop, Andrea and Gonzalez, Michael and Illenčík, Diana and Campos-Castillo, Celeste},
    month = oct,
    year = {2024},
    note = {Publisher: SAGE Publications},
    pages = {5923--5941},
}

@article{de_cicco_millennials_2020,
    title = {Millennials' attitude toward chatbots: an experimental study in a social relationship perspective},
    volume = {48},
    issn = {0959-0552},
    shorttitle = {Millennials' attitude toward chatbots},
    url = {https://doi.org/10.1108/IJRDM-12-2019-0406},
    doi = {10.1108/IJRDM-12-2019-0406},
    number = {11},
    urldate = {2026-01-28},
    journal = {International Journal of Retail \& Distribution Management},
    author = {De Cicco, Roberta and Silva, Susana C. and Alparone, Francesca Romana},
    month = jul,
    year = {2020},
    pages = {1213--1233},
}

@article{skjuve_help_2019,
    title = {Help! {Is} my chatbot falling into the uncanny valley? {An} empirical study of user experience in human-chatbot interaction},
    volume = {15},
    issn = {1795-6889},
    shorttitle = {Help! {Is} my chatbot falling into the uncanny valley?},
    doi = {10.17011/ht/urn.201902201607},
    number = {1},
    journal = {Human Technology},
    author = {Skjuve, Marita and Haugstveit, Ida Maria and Følstad, Asbjørn and Brandtzaeg, Petter Bae},
    year = {2019},
    note = {Place: Finland
Publisher: Agora Center},
    pages = {30--54},
}

@inproceedings{zamora_im_2017,
    address = {New York, NY, USA},
    series = {{HAI} '17},
    title = {I'm {Sorry}, {Dave}, {I}'m {Afraid} {I} {Can}'t {Do} {That}: {Chatbot} {Perception} and {Expectations}},
    isbn = {978-1-4503-5113-3},
    shorttitle = {I'm {Sorry}, {Dave}, {I}'m {Afraid} {I} {Can}'t {Do} {That}},
    url = {https://dl.acm.org/doi/10.1145/3125739.3125766},
    doi = {10.1145/3125739.3125766},
    urldate = {2026-01-28},
    booktitle = {Proceedings of the 5th {International} {Conference} on {Human} {Agent} {Interaction}},
    publisher = {Association for Computing Machinery},
    author = {Zamora, Jennifer},
    month = oct,
    year = {2017},
    pages = {253--260},
}

@misc{IpsosSoMeTrackerQ4232024,
    title = {Ipsos {SoMe}-{Tracker} {Q4}'23 {\textbar} {Ipsos}},
    url = {https://www.ipsos.com/nb-no/ipsos-some-tracker-q423},
    language = {nb-no},
    urldate = {2026-04-01},
    author = {Chernykh, Anastasiia},
    month = may,
    year = {2024},
}

\end{document}